\documentclass[epj]{svjour}
\usepackage{graphics}
\begin{document}
%
\title{World currency exchange rate cross-correlations}
\author{S. Dro\.zd\.z\inst{1,2} \and A.Z. G\'orski \inst{1} \and 
J. Kwapie\'n \inst{1} 
}                     
%
%
\institute{Institute of Nuclear Physics, Polish Academy of Science, PL--31-342
Krak\'ow, Poland
\and Institute of Physics, University of Rzesz\'ow, PL--35-310 Rzesz\'ow,
Poland}
\date{Received: date / Revised version: date}
%
\abstract{
World currency network constitutes one of the most complex structures that 
is associated with the contemporary civilization. On a way towards 
quantifying its characteristics we study the 
cross correlations in changes of the daily foreign exchange rates within 
the basket of 60 currencies in the period December 1998 -- May 2005. Such 
a dynamics turns out to predominantly involve one outstanding eigenvalue 
of the correlation matrix. The magnitude of this eigenvalue depends however 
crucially on which currency is used as a base currency for the remaining ones. 
Most prominent it looks from the perspective of a peripheral currency. 
This largest eigenvalue is seen to systematically decrease and thus
the structure of correlations becomes more heterogeneous, 
when more significant currencies are used as reference. 
An extreme case in this later respect is the USD in the period considered. 
Besides providing further insight into subtle nature of complexity, 
these observations point to a formal procedure that in general 
can be used for practical purposes of measuring the relative 
currencies significance on various time horizons.
} 
\maketitle
%

The financial markets offer an arena to quantitatively view the most 
complex aspects of human activity on the global scale. In terms of volume 
the currency market represents the largest market as its daily 
transactions total trillions of dollars. The related dynamics of the 
Foreign Exchange (FX) market~\cite{ausloos} involves interactions with 
virtually all information around the world and, in particular, with the 
price changes on all other markets. The exchange rates between different 
currencies can thus be well represented by stochastic variables. Within 
the finite time horizons any such rate is therefore likely to lead to a 
nonzero correlation with another~\cite{ivanova,mizuno2,ortega}. Collection 
of the corresponding correlation coefficients can be used to form a 
correlation matrix and the question which emerges is what are the patterns 
that are encoded in such a matrix. For the stock market this issue is by 
now quite well explored in the physics 
literature~\cite{Laloux,Plerou,Kwapien}. For instance, it turns out common 
that the bulk of eigenspectrum of the corresponding correlation matrices 
is largely consistent with predictions of the random matrix theory (RMT). 
Typically only a small fraction of eigenvalues stay sizably apart from the 
bounds prescribed by the RMT. The largest eigenvalue is associated with 
the most collective market component which represents the common factor 
driving the stocks participating in that market. A few smaller eigenvalues 
correspond to similar effects but on the level of various distinct market 
sectors. The logic of the currency market is somewhat different. No 
analogous common factor can for instance be always expected. On the other 
hand there exist currencies -- like at present the USD~\cite{mizuno} -- 
that play a major role in the contemporary world trading system and, on 
the opposite pole, there are those whose global role is marginal. How such 
effects can be quantified remains an issue both of the fundamental as well 
as of practical importance.
       
Our study is based on the daily FX time series~\cite{data} of 60 
world currencies (including platinum, gold and silver, the full list of 
currencies is diplayed in Fig.~3), $x^{(a)}_i(t)$, from the period Dec 
1998 -- May 2005, where the value of $i$-th currency is expressed in terms 
of the base currency $a$. 
The data were
preprocessed in order to remove some numerical artifacts (we did this by
removing day-to-day exchange rate jumps larger than 5$\sigma$, losing in this way
not more than 0.3\% of data points) and to synchronize the gaps related to
non-trading days. For each currency we obtained a time series of
1657 data points.

As usually, the logarithmic daily returns are defined
$G^{(a)}_i(t;\tau) = \ln x^{(a)}_i(t+\tau) - \ln x^{(a)}_i(t)$,
where the return time $\tau$ is also called the time lag. 
For $n$ currencies there are in principle $n\times (n-1)$ exchange rates.
Neglecting friction (transation costs) half of them are simply related to their
inverse counterparts
\begin{equation}
\label{inverse}
 G^{(a)}_i(t;\tau) + G^{(i)}_a(t;\tau) = 0 \ .
\end{equation}
In addition, it can be shown that the remaining exchange rates are not
independent. 
Due to the triangular arbitrage we have additional $(n-2)(n-1)/2$ 
independent constraints called the triangle rules
\begin{equation}
\label{TriangleEffectG}
 G^{(a)}_i(t;\tau) + G^{(i)}_b(t;\tau) + G^{(b)}_a(t;\tau) = 0 \ .
\end{equation}
In effect, we have $(n-1)$ independent time series.

For the present exploratory purpose the above basket of currencies constitutes
a reasonable compromise that represents the global currency network.  
It includes the convertible currencies, from major up to the less liquid
ones, whose price is settled in spot trading, as well as the non-liquid
currencies whose exchange rates are fixed by the central banks. Interestingly,
from the daily time scale perspective the dynamics of fluctuations of all the
corresponding rates within this basket is governed by a similar functional
form. For instance, we have verified explicitely, as one quantitative test, that
the daily return distrbutions for these rates turn out exponential with similar
accuracy. This may - and at least partly even should - originate from the
trading elements that make the triangle rules work, which unavoidably introduces
links between all the existing currencies.
We therefore find it optimal, also for statistical reasons, 
to initiate our analysis with all the 60 currencies taken together.  

In what follows we use the correlation matrix formalism in order to globally 
study the mutual dependencies among changes of the exchange rates within this basket. 
Any of the $n$ currencies can then 
be used as a base currency for the remaining ones.
The correlation matrices are therefore constructed for each 
of those $n$ base currencies separately. 
This results in $n$ matrices of the size $(n-1) \times (n-1)$,  
${\mathbf C}^{(a)} \equiv [C^{(a)}]_{ij}$, 
where $i,j=1,...,n-1$ and $a$ denotes the base currency ($a=1,...,n$).
To construct the correlation matrix one defines the auxiliary matrix
${\mathbf M}^{(a)}$ by normalized returns,
$ g^{(a)}_i(t;\tau) = [ G^{(a)}_i(t;\tau) - 
\langle G^{(a)}_i(t;\tau) \rangle_T ] / \sigma(G^{(a)}_i)$,
where $\sigma(G)$ denotes standard deviation of $G$. 
Taking $(n-1)$ time series 
$\{ g^{(a)}_i(t), g^{(a)}_i(t+\tau), \ldots, g^{(a)}_i(t+(T-1)\tau) \}$
of length $T$ we can built an $(n-1)\times T$ rectangular matrix 
${\mathbf M}^{(a)}$. 
Finally, the symmetric correlation matrix is defined by
\begin{equation}
\label{defCM}
{\mathbf C}^{(a)} \equiv [ C^{(a)} ]_{ij} = \frac{1}{T} \;
{\mathbf M}^{(a)} \;  \widetilde{\mathbf M}^{(a)} \ ,
\end{equation}
where tilde means the matrix transposition. 
Each correlation matrix is determined by its eigenvalues ($\lambda^{(a)}_i$) 
and eigenvectors ($v^a_{ij}\equiv {\mathbf v}^a_i$), 
${\mathbf C}^{(a)} \; {\mathbf v}^{a}_i  =  
\lambda^{(a)}_{(i)} \; {\mathbf v}^a_i$ .
The eigenvalues are enumerated in the monotonic order, $0 \le 
\lambda^{(a)}_{(1)} \le \lambda^{(a)}_{(2)} \le \dots \le 
\lambda^{(a)}_{(n-1)}$. 
The trace of the matrix is always
${\rm Tr} \; {\mathbf C}^{(a)} = n-1$, 
i.e. equal to the number of time series.

In the present study each such matrix is dealt with separately and the 
main focus is put on identification of a potential non-random component.
In general, as compared to a pure random case 
where the entries of the correlation matrix are distributed according to a 
Gaussian centered at zero, the real correlations may correspond either to 
the uniform displacement of such a distribution~\cite{Drozdz1} or to the 
appearance of fatter tails~\cite{Drozdz2}. Both such mechanisms generate 
an effective reduction of the rank of a leading component of the matrix 
under consideration and, as a consequence, a large eigenvalue as compared 
to the remaining ones is expected to be seen. In our case of $n=60$ 
correlation matrices the distribution of their off diagonal entries 
strongly depends on which currency is used as the base currency. Several 
representative cases, including the most extreme ones, are shown in Fig.~1. 
\begin{figure}
\begin{center}
\resizebox{0.29\textwidth}{!}{%
  \includegraphics{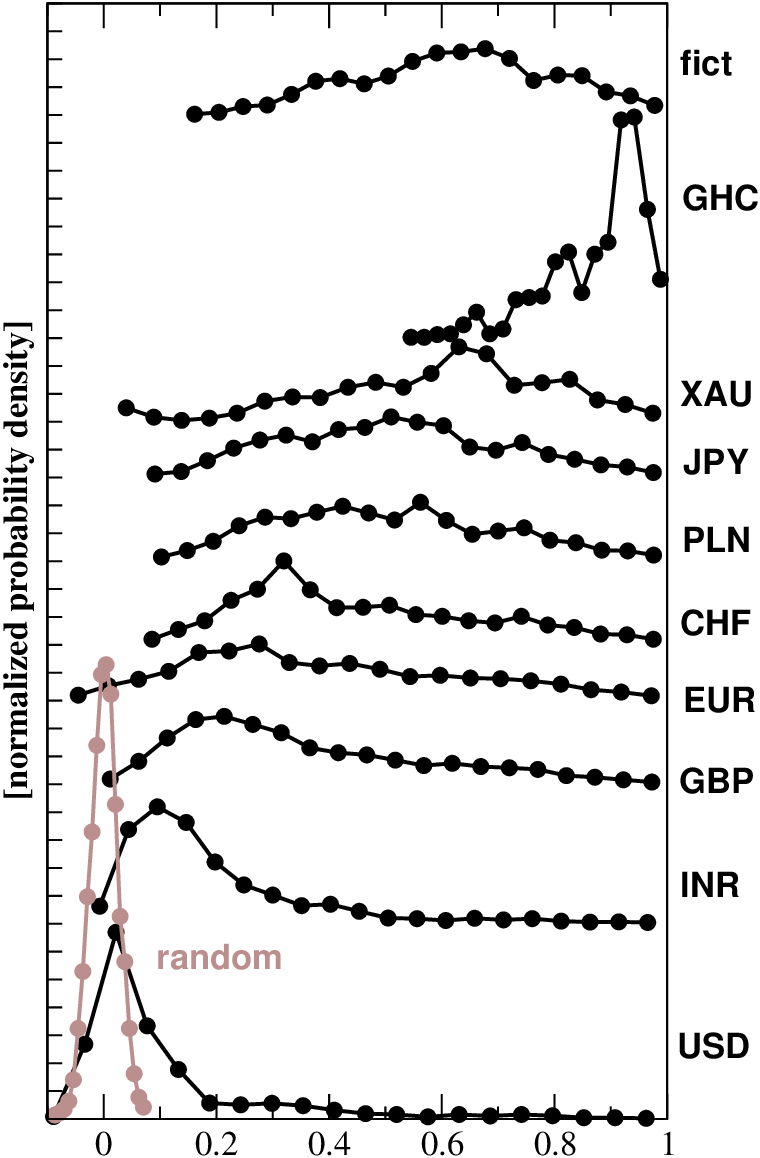}
}
\end{center}
\caption{Distribution of the off diagonal elements of correlation  matrices,
$C^{(a)}_{ij}, i>j$, for selected base currencies ($a$). For the sake of
clarity each curve is appropriately shifted upwards, with vertical scale
preserved. The gray curve corresponds to random (white noise) time series of
returns. The subsequent plots are shifted up by integer number for better
readibility. One tick on the vertical axis corresponds to probability
density difference equal 1.0.}
\label{fig:histograms}       
\end{figure}

First of all, they all differ from a Gaussian centered at 
zero. At first glance somewhat surprisingly however it is the case of the 
USD used as the reference currency that differs least from a 
Gaussian. The shape of the distribution in this case indicates that 
changes among the currencies expressed in the USD are both positively as 
well as negatively correlated. However, there is a visible asymmetry 
towards positive correlations, some of them being close to unity. As we 
move up in Fig.~1, i.e., the other listed currencies are used as 
reference, the FX dynamics becomes more and more positively correlated. 
Both the maximum of the distribution as well as its mean systematically 
move to the larger positive values. In our basket of currencies an extreme
situation one finds when all the currencies are expressed 
in terms of the GHC (Ghanian cedi). In this case almost all pairs of 
exchange rates have correlation coefficients larger than 0.7 and the 
maximum of their distribution is located at around 0.9.  Finally, two 
'null hypotheses' are generated and the corresponding distributions of the 
correlation coefficients also shown in the Fig.~1. One of them, termed
fictitious (fict), is generated in such a way that the USD/fict 
exchange rate time series is represented by a sequence of uncorrelated 
random numbers. The other real currencies are expressed in terms of such
currency by using the relation: currency/fict = currency/USD * 
USD/fict. By construction, the fictituous currency is entirely 
disconnected from the real world economy. Somewhat unexpectedly the 
resulting distribution of the correlation coefficients does not differ 
much from large fraction of cases when the real currencies are used as the 
reference.  In the second (random) case all the time series are replaced by
the sequences of pure uncorrelated random numbers of the same length as the 
original ones. Clearly, this results in the centered at zero
Gaussian distribution of matrix elements.

The most straightforward and compact characteristics 
of the correlation matrix is its eigenspectrum. 
Such full eigenspectra for the cases presented in Fig.~1
are shown in Fig.~2. 
\begin{figure}
\begin{center}
\resizebox{0.32\textwidth}{!}{%
  \includegraphics{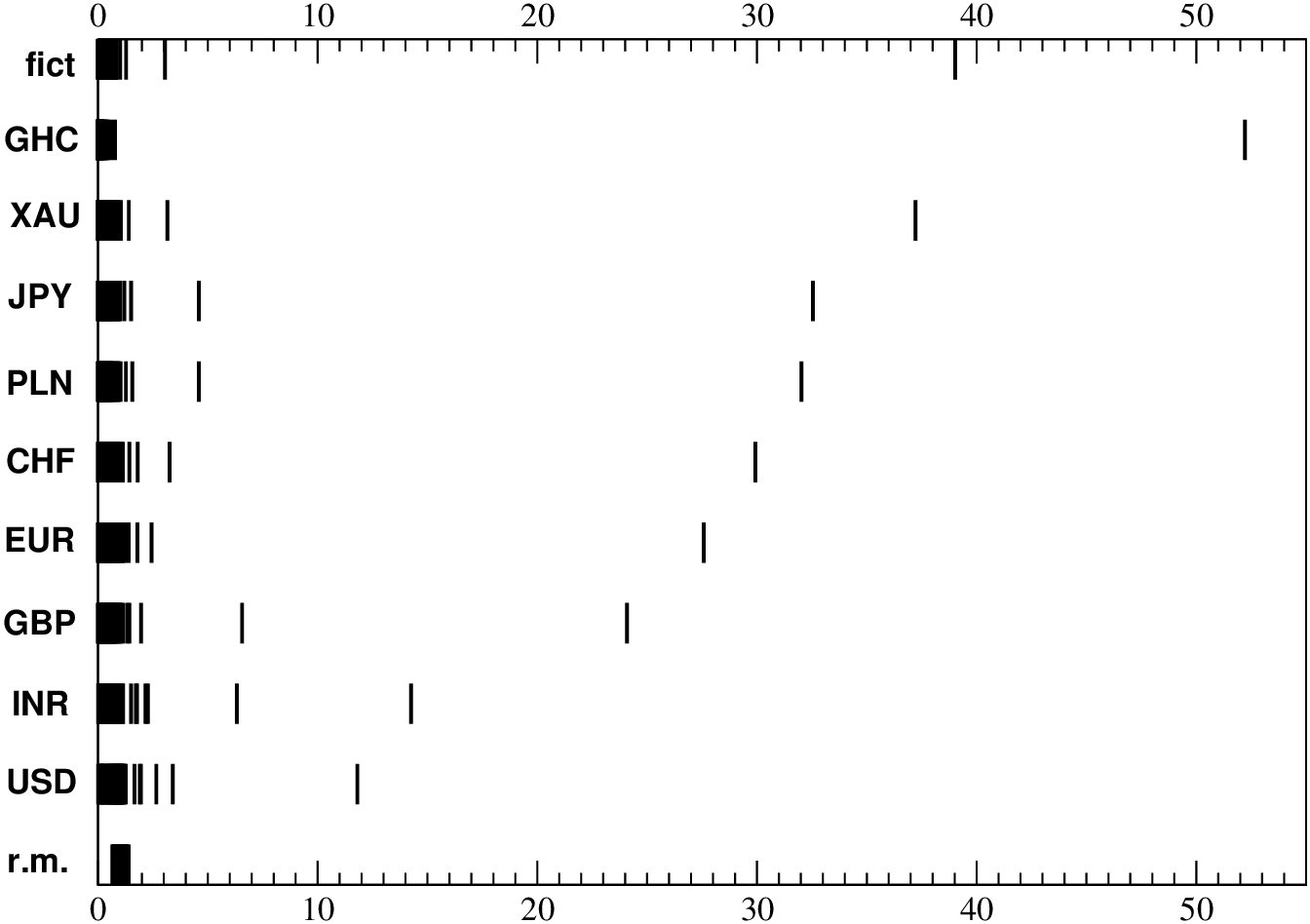}
}
\end{center}
\caption{Eigenspectra ($\lambda^{(a)}_i$) of correlation matrices for the base
currencies as in  Fig.~1. The random matrix eigenspectrum is also displayed
for comparison and denoted as r.m.}
\label{fig:spectra}       
\end{figure}
From our present perspective the most interesting quantity is magnitude of 
the largest eigenvalue, and especially a gap it develops, because it may reflect 
an amount of non-random correlations. The relative locations of these 
largest eigenvalues can easily be seen to go in parallel with the 
differences in distributions of the corresponding matrix elements 
discussed in Fig.~1. Consistently, for the USD we find the smallest value 
while for the GHC the largest. In the later case the correlations are so 
strong that the largest eigenvalue exhausts about 85$\%$ of the total 
matrix trace enslaving~\cite{Haken} the other eigenvalues of this 
correlation matrix to the region close to zero. Several intermediate 
cases, especially when the gold (XAU) is used as the base currency, do not 
differ much from the above described case of a fictitious currency. The 
number of eigenvalues is here too small to perform a fully quantitative 
evaluation of the noise content in the bulk of eigenspectrum by relating 
our results to the predictions of the RMT. In the low lying part of the 
spectra there is some overlap with the case described above as random but 
it seems unlikely that the situation is as extreme as for the stock 
market, even after correcting for the repelling effect of the largest 
eigenvalue. This may originate from the tendency of currencies to exhibit 
correlated clusters at several levels of their 
interactions~\cite{McDonald,FENS06}.
The presence of several (typically 6-8) eigenvalues larger than the RMT 
upper bound can be considered as a manifestation of such effects.

Iterating the above procedure for all the currencies selectively used 
as a base currency one can draw the whole ladder of the corresponding 
largest eigenvalues. The collection of all such eigenvalues for our case
of $n=60$ currencies, plus the case that above is termed a fictitious 
currency, is presented in Fig.~3.
\begin{figure}
\begin{center}
\resizebox{0.45\textwidth}{!}{%
  \includegraphics{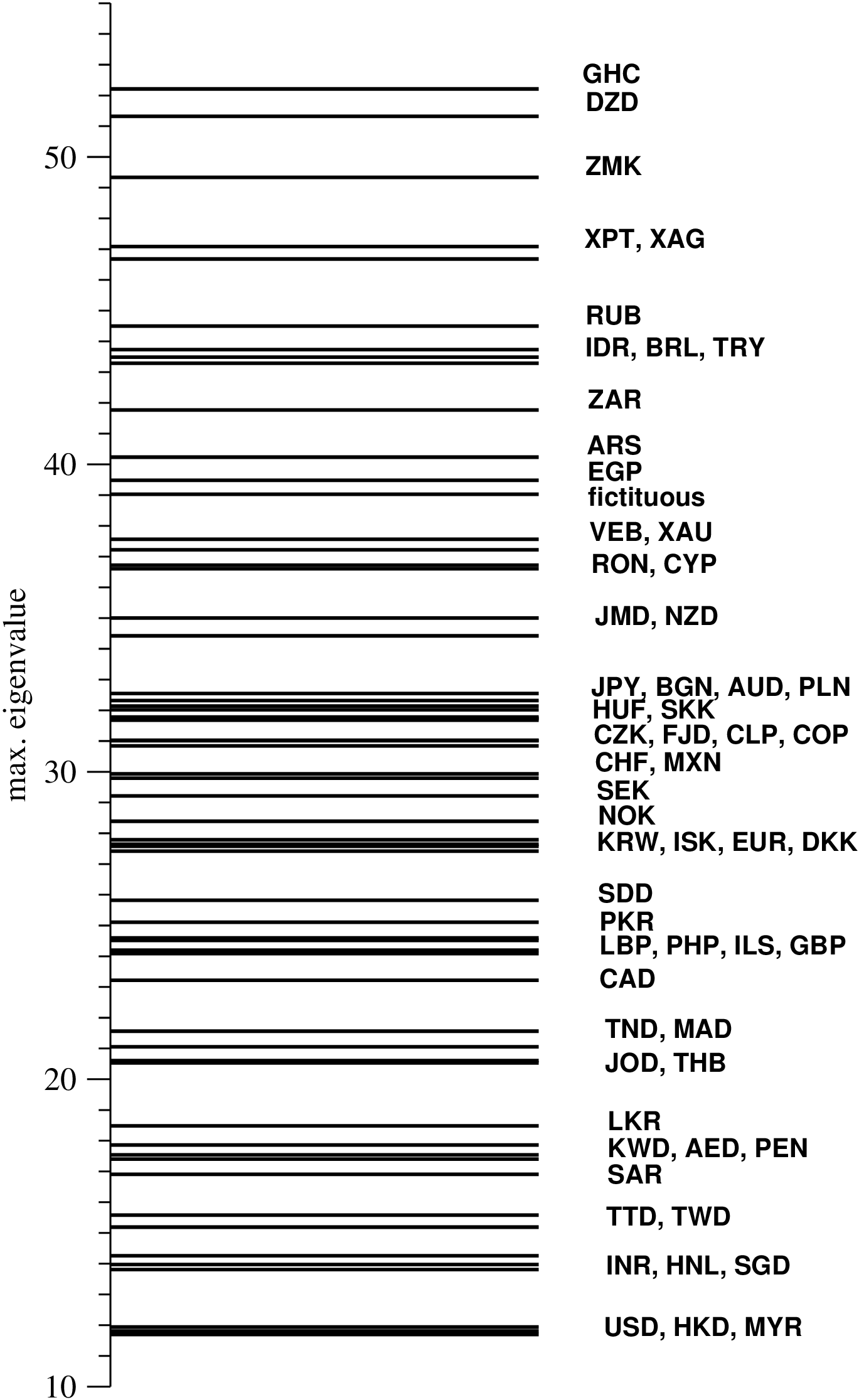}
}
\end{center}
\caption{Maximal eigenvalues ($\lambda^{(a)}_{59}$) for correlation matrices
with all 60 considered currencies taken as the base currency (index $a$), 
including the fictitious (random) currency.}
\label{fig:maxeigenval}       
\end{figure}
The two cases more in detail discussed already above -- of the USD and of 
the GHC -- constitute the lower and the upper bound correspondingly in 
this ladder. Several effects can be discussed based on such a 
representation of the currencies. The one that is especially worth 
drawing attention is that it opens room for assigning a relative 
significance in the world economy to any particular currency. Indeed, it 
is the USD that is most frequently used in the world trading system and 
can be considered the world most influential currency in the period 
studied here. At the same time the exchange rate cross-correlations 
of the other currencies expressed in terms of the USD do involve 
a large eigenvalue which is separated from the rest 
of eigenspectrum by a sizable gap.
The currencies considered do correlate even more 
(larger gap) when expressed in terms of any other selected currency. This 
in particular applies to the fictitious currency which in the ladder of 
Fig.~3 is located relatively high, but still in the region of many 
real currencies. A number of them is represented by even significantly 
larger eigenvalues. 
As it can be inferred already from Fig.~1,
the increase of the largest eigenvalue is 
accompanied by the systematically increasing delocalization - 
almost towards uniformity in case of GHC -
of the corresponding eigenvector components. The magnitude of this largest
eigenvalue can thus be associated with the degree of collectivity.      

A perspective to understand the above results is that the world currency 
network does not pay much attention to changes in value of a peripheral
currency. Majority of the currency exchange rates expressed in that 
particular currency then synchronously adjust in the same direction which
introduces a large reduction of an effective dimensionality of the
corresponding correlation matrix which results in one very large
eigenvalue~\cite{Drozdz3}. The eigenvalues even larger than a peripheral
currency representative, i.e. the fictitious currency eigenvalue,
correspond to those currencies that experience a violent depreciation
(potentially also appreciation) and all the other currencies uniformly
synchronize in reflecting this effect.  On the other hand, changes of     
value of a significant currency -- due to its links to the fundamentals of
the world economy -- may cause a rich diversity of reactions of the other
currencies which in the present formal approach is seen as a sizable
degradation of collectivity. The dynamics in this case is definitely more
complex and the gap between the largest eigenvalue and the rest of the
spectra -- though still pronounced -- is smaller than in any other case
considered, except of course for a somewhat trivial, artificial, totally
random case that develops no gap at all.
This observation provides further arguments in favor of
characteristics that real complexity exhibits when formulated by means of
matrices~\cite{Drozdz3}. 

A note of caution is of course needed as far as blind direct practical
conclusions are to be drawn in the opposite direction. 
This appealing scheme, in which the world's most significant 
currencies used as a base tend to develop relatively small non-random
component while the least influencial currencies correspond to a larger 
collectivity, is blurred in Fig.~3 due to possible strong economic ties 
between different countries or explicit pegs fixing exchange rates of some 
currencies in respect to reference ones. The exchange rates of a pegged 
currency follow the exchange rates of its reference and this obviously 
implies similarity of the eigenspectra of the corresponding correlation 
matrices. For example, the largest eigenvalue for USD is in Fig.~3 
accompanied by the largest eigenvalues for MYR and HKD. In fact, 
significance of neither of the two latter currencies can be compared to 
the significance of American dollar. The coinciding eigenvalues are in 
this case a result of the artificially stablilized HKD/USD rate due to the 
``linked exchange rate'' mechanism in Hong Kong and the MYR/USD peg 
introduced by the Malaysian central bank in response to the Asian crisis 
of 1997. Also in the case of strong economic ties between two countries 
their currencies are expected to behave in a similar way~\cite{ivanova}, 
as it is, for instance, with some Latin American currencies whose values 
fluctuate according to the fluctuations of USD. This effect can also lead 
to a decrease of the largest eigenvalue for the currencies being 
satellites of the world's leading ones like USD or EUR. It should also be 
noted that the analysis based solely on the time series of daily returns 
inevitably neglect such an important economic factor crucial for the 
currency stability as the inflation.

The above reasons indicate that the relation between the currency 
importance and its position in the ladder of Fig.~3 is an 
implication rather than an equivalence. Although the relative currency 
significance affect the magnitude of the associated largest eigenvalue in 
such a way that the more significant currency the smaller the eigenvalue 
gap, the position occupied by a given currency cannot decisively determine 
its significance without some additional knowledge of the country's 
economy and the monetary policy of its central bank. Nevertheless, a 
representation of currencies in the spirit as is illustrated in Fig.~3 
seems to also indicate a practically useful framework to quantitatively 
assign the relative significance of currencies selected 
from their basket at various time horizons.
As a further test it is instructive to make
an analogous analysis in some distinct currencies sectors independently. We
did it for instance for tradeable as well as for non-tradeable currencies
separately and found that the relative locations of the largest eigenvalues
remain similar. 
It thus seems that on daily time scale the mechanism of correlations has some
common elements for all the currencies. This may not be true on the
shorter scales however.

\end{document}